IAC-25-B2,4,2 (99049)

# Commissioning of the TeraNet Optical Ground Station Network

Sascha W. Schediwy[a]*, Aliesha Aden[a], Benjamin Dix-Matthews[a], Alex Frost[a], Amrita Gill[a], David Gozzard[a], Mike Kriele[a], Andrew M. Lance[a], Nicolas G. Maron[a], Ayden McCann[a], Shawn McSorley[a], Lilani Toms-Hardman[a], Shane Walsh[a], Larissa Wiese[a], Graeme Wren[a], Randall Carman[b]

[a] *International Centre for Radio Astronomy Research, The University of Western Australia, Perth WA 6009, Australia*
[b] *Geoscience Australia, Nangetty WA 6522, Australia*

\* Corresponding Author

**Abstract**

TeraNet is a new three-node optical ground station network that has been established in Western Australia. The network is built to support a broad range of space missions operating between low Earth orbit and the Moon, using both conventional and advanced optical technologies developed at the University of Western Australia. It is designed to be spacecraft and mission agnostic, able to be adapted for compatibility with spacecraft using a variety of communication and timing protocols.

The TeraNet network comprises three ground station nodes, each of which are equipped to support direct-detection, bidirectional optical communication with low Earth orbit spacecraft. In addition, each node is focused on the development of a unique advanced optics technology. Specifically:

● TeraNet-1 is a 70 cm aperture optical ground station located on the campus of the University of Western Australia. It uses ultra-sensitive optical detectors and specialised modulation formats to maximise the information recovered from spacecraft at lunar distances.

● TeraNet-2 is a 70 cm aperture, ground station at the Yarragadee Geodetic Observatory 300 km North of Perth. It is equipped with a commercial-grade adaptive optics system for efficient single-mode fibre coupling. This enables high-speed coherent communications and ultra-precise coherent timing and positioning between ground and space.

● TeraNet-3 is a 43 cm aperture mobile optical ground station node built onto the back of a Jeep utility vehicle for rapid, tactical deployment anywhere in the world. It is setup for quantum communication and quantum-assured time transfer and can establish satellite communication links within ten minutes of arriving on site, day or night.

In this paper, we report on the results and outcomes of the TeraNet commissioning campaign, including optical communication links with on-orbit spacecraft, the performance of the TeraNet-2 adaptive optics system, the rapid deployment capability of TeraNet-3, remote network operations of all three nodes, and interoperability tests with other optical ground stations across Australia and New Zealand.

**Keywords:** Optical technology, communications, ground station network,

**Acronyms/Abbreviations**

Adaptive optics (AO), Australasian Optical Ground Station Network (AOGSN), avalanche photodetector (APD), dual polarisation quadrature phase shift keying (DP-QPSK), European Space Agency (ESA), corrected Dall-Kirkham (CDK), Global Navigation Satellite System (GNSS), the International Laser Ranging Service (ILRS), International Terrestrial Reference Frame (ITFR), low-Earth orbit (LEO), New Norcia (NNO), Ritchey-Chrétien (RC), radio frequency (RF), root mean square (RMS), receive (Rx), satellite laser ranging (SLR), TeraNet (TN), transmit (Tx), University of Western Australia (UWA), Yarragadee Geodetic Observatory (YGO).

## 1. Introduction

Free-space optical technology is becoming increasingly important in modern society, particularly for emerging space applications. In these applications, spacecraft are equipped with optical terminals that facilitate optical free space links from the spacecraft back to the ground. Utilisation of free-space optical technology in support of such space applications requires optical ground stations. However, until recently, optical ground stations have been used mostly for experimental development, with limited operation as integrated networks. Maturing these space applications requires the establishment of optical ground station networks. Some particularly relevant space applications include optical communications, optical positioning and timing, and quantum technologies, which will be discussed in more detail in the remainder of this section:

*1.1. Optical Communication*

There is a growing need for high data rate, low-latency space-to-ground communication driven by the increasing demand for Earth observation data and other satellite-provided services such as weather monitoring, real-time traffic monitoring, disaster management, air and sea and remote communications, intercontinental communications, and national security applications. High data rates are also crucial for spacecraft with short-duration orbital transits over key ground stations, and to remove latency bottlenecks for time-critical information, for example in disaster management.

Free-space optical communications has several advantages over radio frequency (RF) communications, including high directionality, greater bandwidth, unregulated spectrum in most jurisdictions, and a smaller physical size enabling use of SmallSats and small ground terminals [1]. These benefits make optical communication valuable to defence, commercial, and scientific applications that require secure, high-speed data transfer over line-of-sight free-space channels.

However, most free-space optical communications experiments by NASA, ESA, DLR, NICT, and others [2-7] have only demonstrated moderate data rates to date, on-par with current RF technology; or they have only operated for a limited-duration. These operational and performance limitations were in part due to operational constraints, including from the deleterious impact of atmospheric turbulence and adverse weather at the ground station sites. Overcoming these issues requires use of advanced communications technology, such as coherent communications [8,9] and adaptive optics [10,11], as well as the establishment of geographically diverse, robust, operational networks of optical ground stations.

*1.2. Optical position and timing*

Satellite laser ranging (SLR) is one of the prime techniques underpinning geodesy, geoscience, and satellite positioning [12], and this is activity is coordinated globally by the International Laser Ranging Service (ILRS) [13]. Satellite positioning is important for operations scheduling, communication coordination, docking, formation flying, and collision avoidance, specifically with space debris [14], which is becoming a critical environmental issue for humanity's continued exploration and utilisation of space.

For geodesy and geoscience, SLR sets the formation of the International Terrestrial Reference Frame (ITRF). The ITRF supports critical applications, including Earth observation from satellites such as Sentinel-6 and GRACE-FO [15], and Global Navigation Satellite Systems (GNSS); however, its current accuracy is five times worse than required by the UN's Committee of Experts on Global Geospatial Information Management. The ITRF can be improved by incorporating new SLR technologies with higher precision and different designs to the other techniques.

Conventional SLR uses pulsed optical signals to measure the position of satellites as a function of time. Greater precision can be achieved by increasing the pulse repetition rate, with typical modern system operating at kHz rates. However, new coherent techniques developed by the TeraNet team at the University of Western Australia (UWA), such as optical Doppler orbitography, are expected to improve the measured precision of satellite orbital parameters by several orders of magnitude [16]. Several proof-of-principle experiments have been conducted to this end, for example, a recent experiment using a hexacopter as a proxy satellite, demonstrated a residual range and range rate precision of 0.1 μm and 100 μm/s, respectively [17].

One limitation with such coherent techniques is that typically, absolute values cannot be determined, only changes in values can be meaured. For example, changes in range rate can be measured, but not absolute range. However, recent laboratory experiments have extended this technique to use mixing of multiple optical signals to demonstrate absolute ranging with a precision of 0.2 μm, which is potentially sufficient to resolve individual optical fringes [18].

*1.3. Quantum technologies*

Quantum technologies can be used for quantum-assured timing and quantum secure communications [19]. Quantum secure communication, specifically discreet variable quantum key distribution, has been demonstrated over space links, most notably between China and Austria via the Micius satellite [20], and between China and South Africa via the Jinan-1 micro-satellite [21]. These quantum technologies provide the means to secure communications utilising the physical fundamental properties so bring benefit for national security, and other applications such as banking.

Quantum-assured timing uses similar technology to secure the transfer of absolute time. In this application, entangled photons are transmitted from one site to another with the individual photon arrival times being keyed to indicate absolute time stamps. Interference of these photons can be detected by the user at the remote end, thereby assuring that correctly received timing signals have not been tampered with and therefore the received timing signals are accurate.

*1.4. Optical ground station networks*

Optical communications, optical positioning and timing (with the exception of SLR), and quantum technologies are still being developed, with researchers requiring optical ground stations to validate and space-qualify new technology developments. While there is some technology overlap between these optical technologies, each still carries significantly different requirements for the ground station. This includes optical wavelength, link budget (and therefore ground station aperture size), detector technology (including single- or multi-mode fibre coupling, or free space detection), and backend photonics and digital technologies. This therefore drives the need to have multiple ground stations that are available for testing, located in different parts of the world, that are widely compatible with each of these free space optical technologies.

Optical ground station development has been active for decades, most notably in the US, Japan, and Germany, but it has only been in recent years that larger number of research ground stations, as well as commercially operated ground stations, have been established around the world. However, even now, most of these operate to support specific research or commercial applications. One notable exception is the European Optical Nucleus Network [22], which included a test campaign that spanned several optical ground stations across Europe.

We have developed new Australian capability in this technology sector by developing TeraNet, a three-node optical ground station network established in Western Australia [23]. The network has been built to support a broad range of space missions operating between low Earth orbit (LEO) and the Moon, using both conventional and advanced optical technologies developed at the University of Western Australia (UWA). TeraNet is designed to be spacecraft agnostic to support a number of different optical communications, optical positioning and timing, and quantum technologies.

The location of TeraNet within Western Australia brings significant geographical advantage by covering a unique and critical latitude and longitude of the Earth's surface that cannot be covered by other jurisdictions. Western Australia's longitude enables first contact with Earth observation and communication satellites (which are often in Sun synchronous orbits), after they have just passed over the Asian landmass. This allows for the rapid downlink of critical information relating to around half the world's population. Western Australia's latitude is shared only with the remainder of Australasia, South America and southern Africa.

TeraNet is also part of the Australasian Optical Ground Station Network (AOGSN) [24,25], thereby demonstrating compatibility and networking potential of optical ground stations across Australia and New Zealand.

## 2. Functional Description

### 2.1. Overview

As described in the previous section, the TeraNet network comprises three ground station nodes, each of which is equipped to support direct-detection, bidirectional optical communication with low Earth orbit spacecraft. In addition, each node is focused on developing advanced optical technologies described in the previous section. Each TeraNet node is described in more detail below:

● TeraNet-1 (TN-1): a 70 cm aperture optical ground station located on the campus of UWA in Perth. It uses ultra-sensitive detectors and specialised modulation formats to maximise the information recovered from spacecraft at lunar distances.

● TeraNet-2 (TN-2): a 70 cm aperture optical ground station node at the Yarragadee Geodetic Observatory, 300 km North of Perth. It is equipped with a commercial-grade adaptive optics system for efficient single-mode fibre coupling, which enables high-speed coherent communications and ultra-precise coherent timing and positioning.

● TeraNet-3 (TN-3): a 43 cm aperture transportable optical ground station, built onto the back of Jeep utility vehicle enabling rapid, tactical deployment. It is setup for quantum communication and quantum-assured time transfer and can establish satellite communication links within ten minutes of arriving on site, day or night.

An overview of the TeraNet network functional design and working principle is illustrated Fig. 1. An external user requests spacecraft passes via an operation centre. The operation centre controls the three optical ground station nodes in the network: TN-1, TN2 and TN3. Individual nodes can establish and bidirectional optical link with a spacecraft.

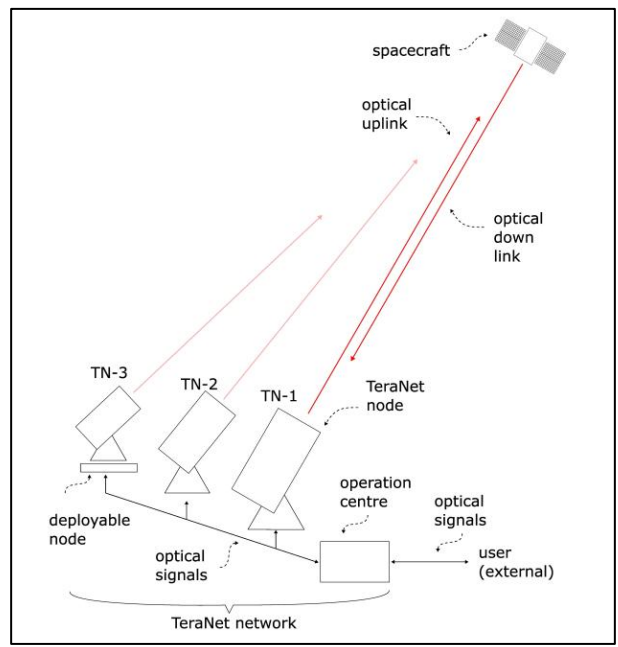

Fig. 1. Illustrates an overview of the functional design and working principle of the TeraNet network.

A map of ground station sites and TN-3 remote deployment locations is shown in Fig. 2. The geographical position of the TeraNet OGS nodes within WA provides resilient access of spacecraft to the TeraNet network when cloud coverage or other adverse weather conditions affect individual nodes.

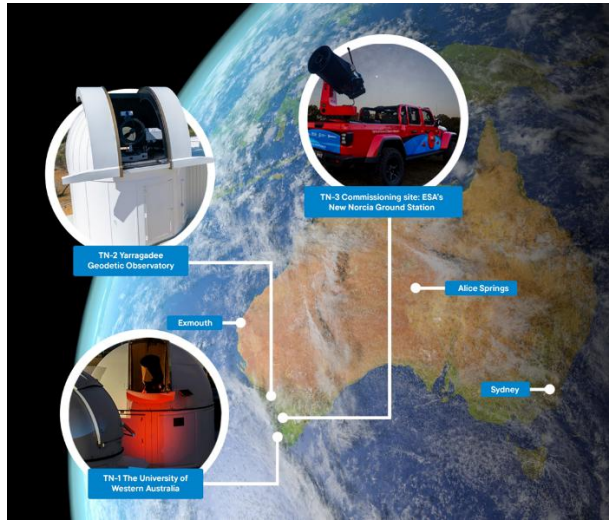

Fig. 2. Map of TeraNet ground station sites: TN-1 Perth; TN-2 Yarragadee Geodetic Observatory; TN-3 ESA commissioning site at New Norcia Deep Space Ground Station; and three planned TN-3 remote deployment locations in Exmouth, Alice Springs, and Sydney.

### 2.2. System Description

Fig. 4, on the following page, provides a simplified functional block diagram of the TeraNet system as a whole; columns show the node configurations; rows illustrate the high-level functional subsystems for each TeraNet node. Note: both the TN-1 and the TN-2 telescopes, each have two Nasmyth optical ports, hence they can each support two distinct optics systems, while TN-3 only has a single Cassegrain optical port, and so only supports one optics system.

## 3. Ground Station Node Details

### 3.1. TeraNet-1 Description

The TN-1 ground station is located on the roof of the five-storey Physics building at the Perth campus of the UWA, as shown in Fig. 3. TN-1 was completed in November 2021. It has since been used to investigate the suitability of laser communications with spacecraft in deep space [26], and it participated in the downlink campaign with the German *Flying Laptop* satellite in July 2024 [27]. It was upgraded for two-way laser communication in July 2025.

The optical ground station's proximity to the UWA technical team, and its location directly over our laser laboratory (enabling direct fibre access), makes it ideal for prototyping and rapid research and development.

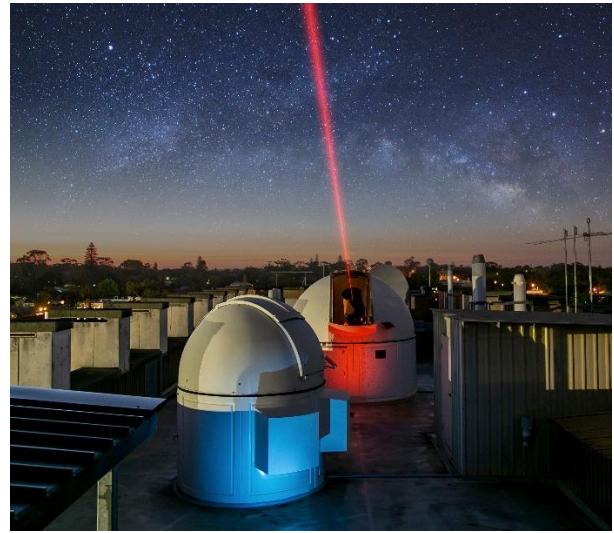

Fig. 3. Artistic representation of TN-1 at UWA in Perth.

The ground station site is connected to a high-speed optical fibre connection and 320 km of dark fibre provided by AARNet (Australia's national research and education network), making it ideal for testing novel optical ranging and timing technologies. Its location at the centre of a major urban area and its proximity to an international airport, empowers daily operations within the sort of environment and conditions that will be faced by many other operational ground stations, which will not all have the luxury of being placed in ideal locations.

#### 3.1.1. TN-1 Telescope Systems

TN-1 is based on a 70 cm aperture, f/6.5, PlaneWave Instruments CDK700 telescope, housed in a 3.5 m Sirius Observatories telescope dome. The corrector lens has been removed from the optics system #1 port in this corrected Dall-Kirkham (CDK) telescope, as off-axis imaging is not required. The telescope includes two Nasmyth ports, allowing for the attachment of two optics systems that remain rotationally-invariant with the main optical axis of the telescope (see Fig. 5). The motors and control system of the telescope dome have been upgraded to track the movement of the telescope at the angular rates required for LEO satellites and to rapidly slew to new targets.

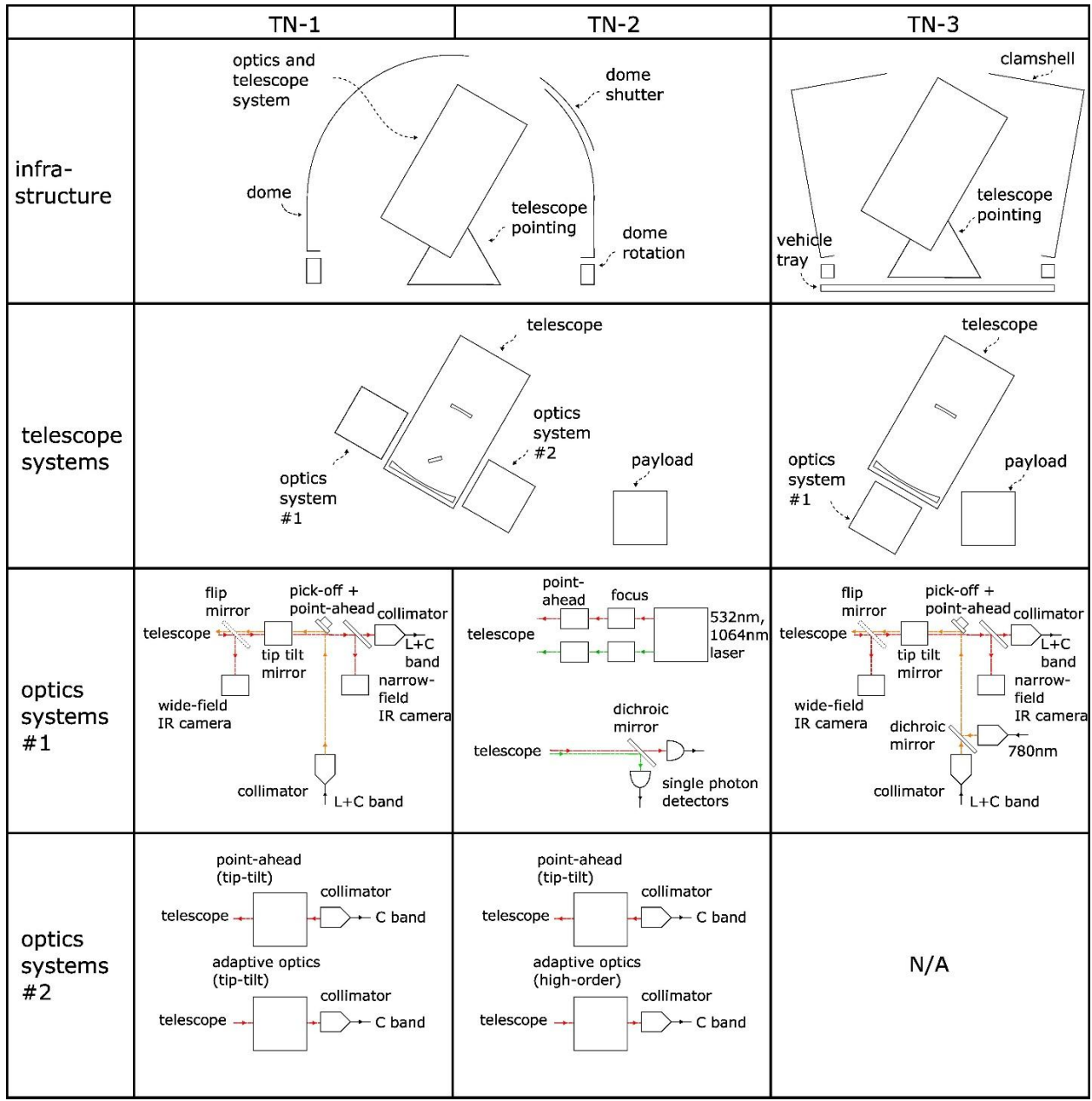

Fig. 4. A simplified functional block diagram of the TeraNet system. The three columns show functional composition of the three TeraNet nodes. The rows show the high-level functional subsystems.

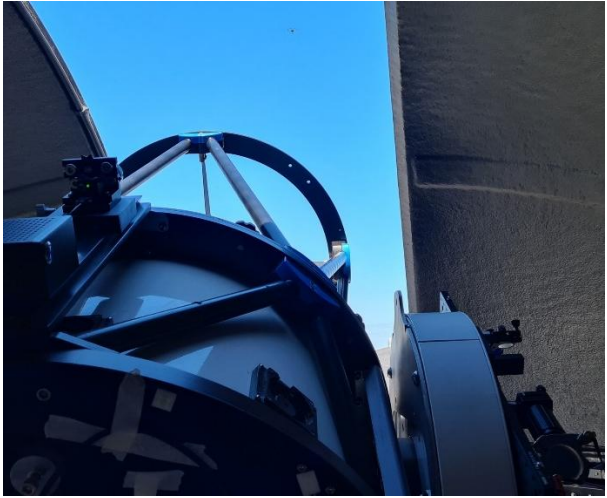

Fig. 5. The TN-1 CDK700 telescope showing a part of optics system #1 on one of the TN-1's Nasmyth ports.

*3.1.2. TN-1 Optics Systems*

TN-1 includes two optics systems (see Fig. 4). Optics system #1 uses tip-tilt adaptive optics on the receive (Rx) path and launches the transmit (Tx) beam through a ~20 cm-diameter unobstructed quadrant of the primary aperture. This results in optimal beam quality and lower divergence, at the expense of ~1 dB Rx loss from the pick-off mirror. The Tx path includes a steerable mirror to adjust the point-ahead angle.

Optics system #2 also incorporates an Rx path with tip-tilt adaptive optics on the primary 70 cm aperture, but has the Tx path through a separate 15 cm, Ritchey-Chrétien (RC) telescope. This provides lower Tx beam quality but better isolation between Tx and Rx beams.

*3.1.3. TN-1 Payload Technologies*

Payload #1 is optimised for conventional 'direct detection' optical communications with spacecraft in LEO. Theses optical communications architectures utilise only amplitude modulation of the optical-frequency carrier signal and are therefore limited to relatively low data rates (~Gb/s); however, the signal can be directly demodulated on a photodetector without any additional photonic circuits.

These direct detection technologies utilise relatively slow photodetectors with proportionally large sensor sizes, thereby allowing for convenient multi-mode optical fibre (or free space) optical signal direction onto the detector sensor. Given the relatively strong optical signal strength expected from LEO spacecraft, the extra loss and beam profile disturbance from the Tx pick-off mirror does not limit the operational performance. The payload includes a Tx point-ahead capability through a separate tip-tilt mirror, for efficient power delivery to LEO satellites that transit at angular rates up to 1 deg/s.

Payload #2 is designed for advanced 'high photon efficiency' optical communications with spacecraft at lunar distances. At these distances, the Rx signals are extremely weak, so information-per-bit is maximised using pulse-position modulation [26]. In addition, ultra-sensitive detectors, including superconducting nanowire single-photon detectors, can be employed to maximize the information recovered. The separate Tx aperture helps with isolating the high-power transmit from the extremely weak Rx signals to stop the sensitive detectors from saturating.

*3.2. TeraNet-2 Description*

TN-2 (see Fig. 6) achieved first light in September 2025. It is hosted at the Geoscience Australia Yarragadee Geodetic Observatory (YGO) within the Mingenew Space Precinct, around 320 km North of Perth. It is co-located with NASA's MOBLAS-5 SLR facility and numerous private and national satellite communication facilities. TN-2 benefits from an ideal location with exceptionally low annual cloud cover, a restricted airspace no-fly zone, grid-power, and access to redundant AARNet high-speed optical fibre connections, thereby ensuring robust and reliable operations.

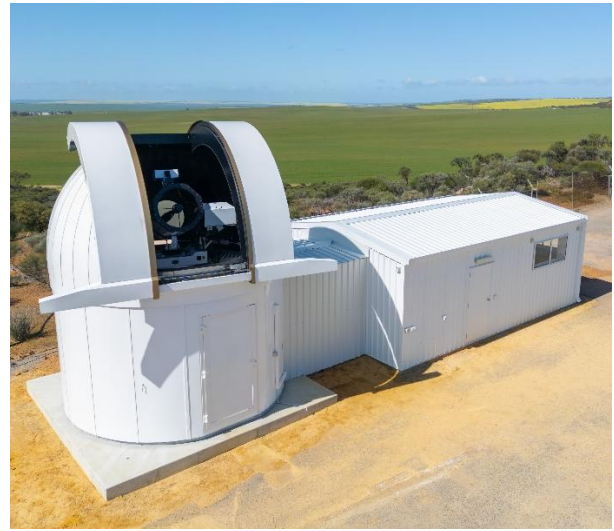

Fig. 6. Photo of TN-2 facility at the Yarragadee Geodetic Observatory.

*3.2.1. TN-2 Telescope Systems*

The heart of TN-2 is a 70 cm aperture, PlaneWave Instruments, f/12 ratio RC700 telescope. The telescope features a <30 % central obstruction and gold-coated mirror surface to optimise for infrared laser signal throughout. The telescope also includes two Nasmyth ports allowing for the attachment of two optics systems, and a Coudé path to laser laboratory in the connected annex building.

The telescope is housed in a two-storey, 4.6 m diameter Gambato planetarium dome with a connected 45 $m^2$ annex building. The annex building "Wren's Nest" includes office facilities, a storage and workshop area, and a secured laser laboratory with an optics table. Both the telescope pier and optics table are on independent foundations to the souring buildings.

### 3.2.2. TN-2 Optics Systems

As shown in Fig. 4, TN-2 also includes two optics systems. Optics system #1 comprises two static co-aligned, laser apertures, one optimised for 532 nm light and the other for 1064 nm. These have relatively small apertures to intentionally broaden the divergence of the transmission beams. They are also unobstructed, making them suitable for high-power transmission, to counteract the low power density resulting from the higher beam divergence. The Rx path consist of a high time-resolution, large sensor size, free space photodetector, and associated phase and amplitude calibration optics.

Optics system #2 is centred around a HartSCI ClearStar LaserComm adaptive optics (AO) system optimised for operations in the 1530-1650 nm range (see Fig. 7). The Rx portion of the AO system combines a 10×10 element Shack-Hartman wavefront sensor with a 97-segment deformable mirror for efficient single-mode fibre coupling. This is combined with a built-in pointing/acquisition camera and calibration source.

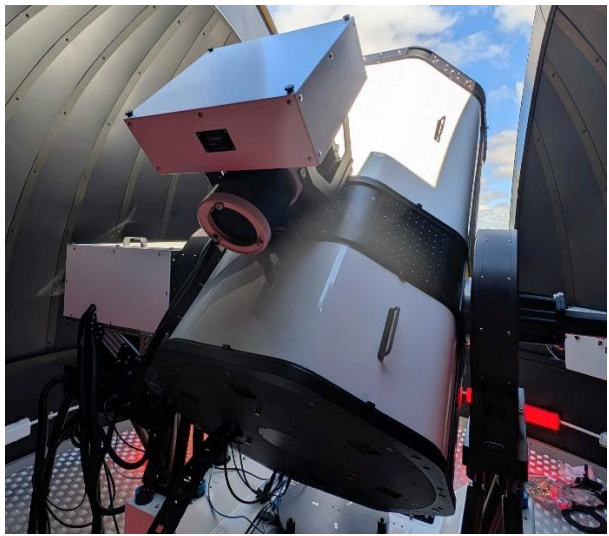

Fig. 7. TN-2 optics system #2.

The AO Tx comprises an uplink compensation system with an 85 mm beam, that includes a steering mirror to account for the variable point-ahead angle, supporting uplink laser power up to 10 W.

### 3.2.3. TN-2 Payload Technologies

Optics system #1 supports a state-of-the-art, kHz-class, dual-wavelength SLR capability. This is driven by a picosecond laser with two controllable outputs at 532 nm and 1064 nm, with timing jitter of less than 300 ns. The Tx beam for the 532 nm is sufficiently defocalised to allow for eye-safe ranging to the International Space Station to support ESA's Atomic Clock Ensemble in Space mission [28]. The divergence of the 1064 nm Tx beam is set to optimise the maximum Rx power and target acquisition time.

The AO system on optics system #2 supports a range of cutting-edge advanced technologies that rely on precise optical phase control or measurement, including ultra-high-speed coherent communications, coherent optical positioning and timing, and quantum technologies. Initial research into ultra-high-speed coherent communications [8,9] will involve a 28 Gbaud dual polarisation quadrature phase shift keying (DP-QPSK) digital coherent optics module, with net line rate of 118.8 Gbps. Coherent optical positioning and timing [16,17,18] and chronometric geodesy [29,30] will be realised using UWA's velociRAPTOR and digital multi-wavelength optical absolute ranging systems. Finally, quantum technologies, such as quantum-assured timing [19] and quantum imaging [ 31 ] can utilise the combination of the optics system #2 and the laser laboratory accessed through the telescope's Coudé path.

### 3.3. TeraNet-3 Description

TN-3 is a fully mobile 43-cm optical ground station mounted on a Jeep utility vehicle that offers unparalleled versatility for optical communication operations (Fig. 8).

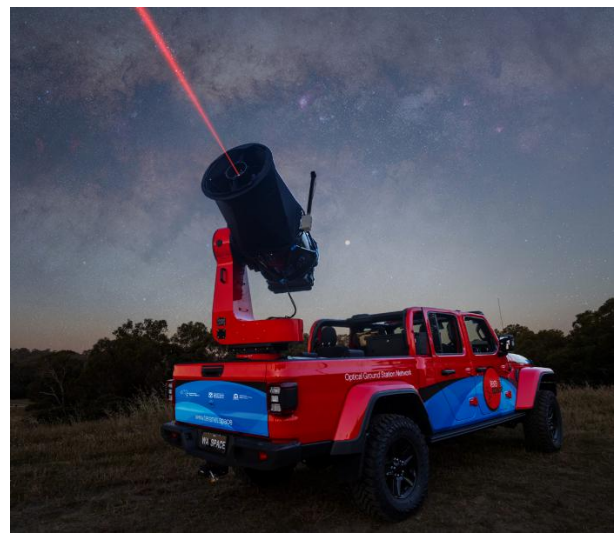

Fig. 8. Artistic representation of TN-3.

This compact and self-contained system does not require external power or levelling systems, enabling rapid deployment in diverse and remote environments. It can be operational in under 10 minutes, during day or night, ideal for dynamic and real-time connectivity.

TN-3 was first used to receive laser signals from the German *Flying Laptop* satellite in July 2024, and it has since conducted downlink campaigns at various sites across Perth, and at its primary commissioning site, the ESA' New Norcia (NNO) Deep Space Ground Station. Future downlink campaign deployment locations include Exmouth in WA and central Sydney in NSW.

### 3.3.1. TN-3 Telescope Systems

The TN-3 telescope system comprises a 43 cm, f/6.8, PlaneWave Instruments CDK17 (with corrector plate removed) telescope on an L-500 mount. This is attached to a mechanical raising/lowering system that can fold down onto a steel pallet, which can be easily removed with a pallet jack (see Fig. 9).

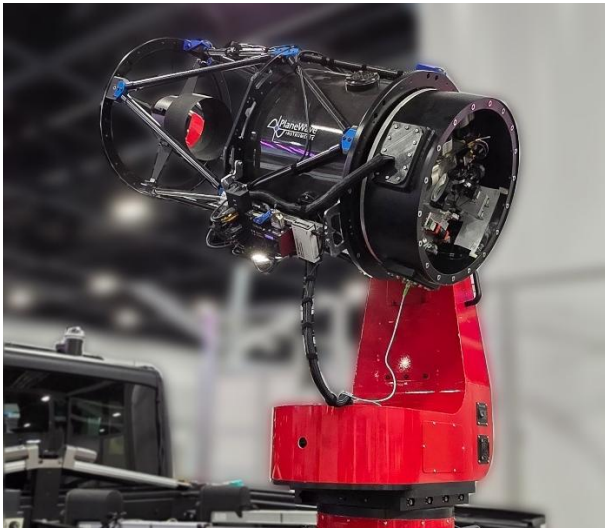

Fig. 9. Photo of the TN-3 telescope system.

The pallet is mounted in the bed of a Jeep Gladiator utility vehicle, where is shielded from the external environment when not in use by a mechanical clamshell opening/closing system. A 12U, 19" equipment rack is mounted behind the vehicle's driver position. The ground station can be operated fully autonomously, or it can be operated by two users seated in the vehicle during operations, one to manually control or supervise the telescope system and the second the payload.

### 3.3.2. TN-3 Optics Systems

The CDK17 has a single Cassegrain port and this is utilised for two optics systems. Optics system #1 is similar that that of TN-1, in that it uses tip-tilt adaptive optics on the Rx path, and it launches the Tx beam through a ~11 cm-diameter unobstructed quadrant of the primary aperture. However, it is optimised for operations in the 1530-1650 nm band through use of a dichroic splitter. The Tx path includes a steerable mirror to accommodate adjustments in point-ahead angle.

Optics system #2 uses a common receive path as optics system #1, but as it is optimised for operations around 800 nm, with the output being directed to a different fibre coupler. For this system, the Tx is launched via a separate smaller telescope that is underslung on the primary optical tube assembly.

### 3.3.3. TN-3 Payload Technologies

Payload #1 is optimised for direct detection optical communications with spacecraft in LEO. The optical signals are coupled into a multi-mode fibre on the telescope and routed to a receiver located in the equipment rack within the vehicle cabin and then digitised and demodulated. Payload #2 is designed for transmission of 800 nm entangled photons for quantum assured timing applications. A camera is used to receive the optical beacon at 960 nm from the satellite.

## 4. TeraNet commissioning campaign outcomes

Commissioning of the TeraNet optical ground station network started in 2024 with preliminary work involving TN-1 and TN-3, and it has since moved onto full system commissioning with the conclusion of TN-2 construction in September 2025. This commissioning phase is scheduled to be completed by early 2026. Past commissioning phase activities have included communication link campaigns with on-orbit spacecraft; TN-2 adaptive optics system performance evaluation, both in a laboratory setting over a horizonal free-space link and on-sky when mounted on the telescope; demonstration of the rapid deployment capability and performance of the TN-3 Jeep; the remote operations of the three nodes as a network; and participation in interoperability and network availability tests with three other AOGSN ground station nodes.

### 4.1. Communication links with on-orbit spacecraft

All three optical ground station nodes of the TeraNet network have engaged in on-orbit communications links campaigns, starting with a preliminary *Flying Laptop* downlink campaign in July 2024, using TN-1 and TN-3. Since then, the optics systems for both TN-1 and TN-3 have been completed to include appropriate transmit paths, and the full duplex optics system (Tx and Rx) for TN-2 has also been completed.

For the *Flying Laptop* campaigns, the optics systems connected to a receive module that comprises a commercial variable gain avalanche photodetector (APD) and a digitiser (AlazarTech ATS9360). The InGaAs APD, used to conduct optical-to-electrical conversion of the downlink beam, has an active area of 0.2 mm, a bandwidth of 400 MHz, and a saturation power of −16.6 dBm. The digitiser, used to record signal acquisitions, has an input range of ± 400 mV. It is set to oversample the 39 Mbps 8-bit repeating signal at a sample rate of 500 MS/s. Each acquisition contains $10^6$ samples and is therefore 2 ms long. On MATLAB, these acquisitions can be analysed using various digital signal processing techniques, assessing its performance in terms of signal-to-noise ratios, bit-error rates and corresponding quality (Q)-factor values amongst other performance parameters.

### 4.2. TN-2 adaptive optics system performance

The TN-2 AO system control loop can operate at up to a 1 kHz speed, depending on the source brightness. The 10×10 Shack-Hartmann and deformable mirror can reliably correct up to 94 spatial modes of the beam wavefront. HartSCI have dedicated significant effort to developing alignment and calibration routines, making the AO system extremely robust and tuneable. Notable features include gross focus offloading to the RC700 secondary mirror, offloading of accumulated tip/tilt errors to the telescope mount, and an automated periscope for co-locating the optical axes of the telescope and AO system. Further, inbuilt alignment sources allow influence matrices to be recalculated at any point. The target wavefront can also be optimised for single-mode fibre coupling through a gradient-ascent routine. When on-sky, real-time adjustment of the compensation mode number and the control loop bandwidth is possible.

Initial testing using stellar sources, demonstrated that even at power levels that are two to three orders-of-magnitude lower than would be expected to be received for typical LEO satellites, the AO system has sufficient sensitivity and bandwidth to concentrate the majority of the receive power to within the size of a single mode fibre core; see Fig. 10. The residual astigmatism is due to an error on the AO deformable mirror that will be resolved on the next internal calibration.

The performance of the AO system is expected to improve when operating at nominal receive power levels, and this result is a very promising indication that this AO system will also perform well when receiving signal from mid Earth orbiting satellites.

### 4.3. TN-3 rapid deployment capability

The TN-3 Jeep is the world's first rapidly deployable optical ground station, in that it can arrive at an unprepared site, be mechanically setup, and be on-sky with an appropriate telescope pointing model in under 10 minutes, during day or night [32]. Some of the key factors that contribute to this rapid deployment capability, is the that 1). the primary 43 cm telescope is built onto the back of a nimble, go-anywhere Jeep Gladiator utility vehicle; 2). the Jeep features a mechanised opening canopy and mechanised telescope raising platform; 3) it includes an onboard inertial measurement unit and ancillary sensors to allow the vehicle to be parked in arbitrary orientation on non-even surfaces; 4) it uses an innovate rapid and precise star-based, infra-red calibration technique that can work during the day and night; and 5) it incorporates a fast tip-tilt adaptive optics system to remove any residual movement of the vehicle (e.g. from wind buffeting or personnel movement) thereby eliminating the need for rigid stabilising legs.

These factors have come together to result in a system that has a demonstrated on-sky root mean square (RMS) pointing errors of 3.54 arcseconds (17.2 µrad) and a mean RMS closed-loop tracking error of 1.14 arcseconds (5.54 µrad) across multiple LEO satellite passes [32]. More recently, the system was deployed for an initial remote satellite link operations test campaign at ESA's NNO Deep Space Ground Station in September 2025, see Fig. 11.

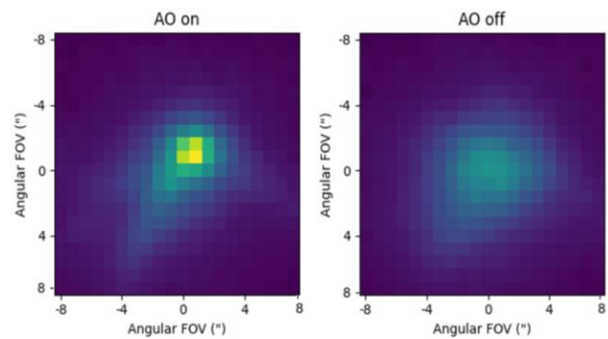

Fig. 10. Heat map showing the distribution of power received with the AO system on (left panel) and off (right panel) for the star Antares. Each pixel is approximately 10 µm wide, matching size of a single mode fibre core.

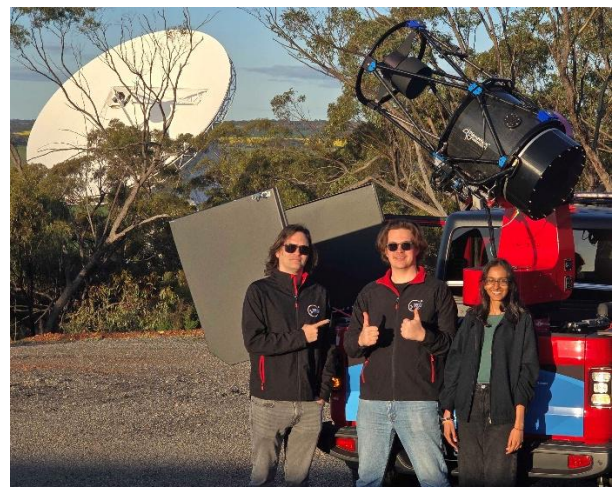

Fig. 11. TN-3 initial remote satellite link operations test campaign at ESA's NNO Deep Space Ground Station.

*4.4. Remote network operations of the three nodes*

Each of the three TeraNet nodes is capable of being operated remotely. Fig. 12 is a photo of the ground station control centre that was setup at UWA in November 2011 for this purpose. The robustness of these remote operations, and any present limitations, will be validated as part of the remaining TeraNet commissioning campaign.

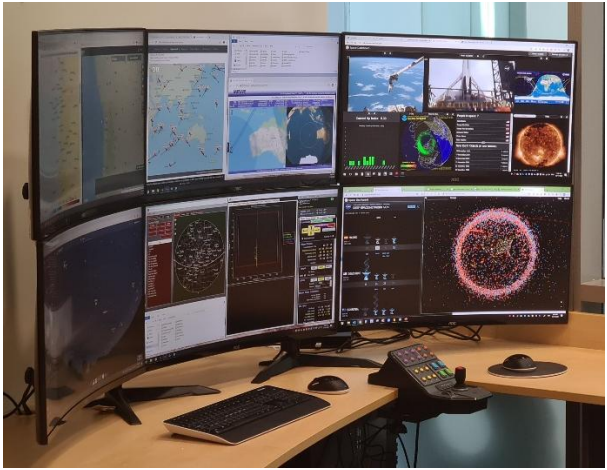

Fig. 12. The optical ground station network operations centre, established at UWA in November 2011.

Both the two fixed ground station sites at UWA and Yarragadee Geodetic Observatory, and the primary commissioning site for TN-3 at NNO, include high-speed AARNet optical fibre links back to the optical ground station network operations centre, established at UWA. The main computers for each of the three stations can be controlled from the network operations centre, or from an ad-hoc facilities anywhere with a connection to the internet. In addition, the operations centre has access to the stations' all-sky camera, remote monitor webcams, weather station data, and other ancillary data.

The concept of operations involves the network user logging into the node of choice, confirms suitable weather conditions using a combination of node-specific sensors (all sky camera and weather data) and external weather information, confirms no physical blockages using the webcams, and commands the dome to open. In the case of TN-3, the canopy opens, and the telescope mechanism is raised. All three nodes are located within secured sites with no public access, and TN-3 has a flashing warning light that is engaged when the canopy and raising mechanism is in motion. The user can then operate the station as required, and then close-up and shut-down the station at the termination of their run.

*4.5. Interoperability tests with the AOGSN*

TeraNet participated in interoperability and network availability tests with three other AOGSN ground station nodes in July 2026 [25]. All four nodes were tasked to track every nighttime pass of five bright LEO satellites for a period of five consecutive nights. An online dashboard was created to display the upcoming satellite pass information and to show the relevant all-sky camera, weather, and telescope operations data for each site (see Fig. 2. of [25]).

Each site could then report on the status of the pass, including node availability, weather go/no go status, and whether the satellite tracking was attempted. This week-long campaign produced an analysis of network availability, taking each of these factors into account. The experimental methods, results, and a detailed discussion of this campaign is presented in [25].

## 5. Discussion and Conclusion

In this paper, we presented an overview of TeraNet, a three-node optical ground station network operating in Australia. The network is built to support a broad range of space missions operating between low Earth orbit and the Moon, using both conventional and advanced optical technologies. It is designed to be spacecraft agnostic, able to be adapted for compatibility with spacecraft using a variety of communication and timing protocols. Each node is equipped to support conventional direct-detection, bidirectional optical communication with low Earth orbit spacecraft; however, each node is also optimised to focus on the development of a specific advanced optics technology.

The paper also reported on the results and outcomes of the TeraNet commissioning campaign, including on optical communication links with on-orbit spacecraft with each of the nodes, the performance of the TN-2 adaptive optics system, the rapid deployment capability of TN-3, remote network operations of all three nodes, and interoperability tests with other optical ground stations across Australia and New Zealand. The commissioning and full performance validation of the TeraNet network will be completed in the first half of 2026, after which time, the network is expected to transition to operations.

With free-space optical technology becoming increasingly important for emerging space applications, the establishment of versatile optical ground station networks, is very timely. Given TeraNet's ability to support multiple user application spanning across optical communications, optical positioning and timing, and quantum technologies, and its location in highly-sought-after and well-suited global region for optical ground stations, makes TeraNet a critical asset for enhancing Australia's sovereign capability and contributing to globally significant developments.

**Acknowledgements**

This work is supported by our project partners Thales Australia, Thales Alenia Space, and Geoscience Australia, and our founding project collaborators DLR, ESA, CNES, AARNet, and Fugro SpAARC. TeraNet receives funding from the Australian Space Agency, the Government of Western Australia, and the University of Western Australia.